\documentclass[aps,prl,reprint,tightenlines,superscriptaddress,amsmath,amssymb]{revtex4-1} 

\usepackage[latin1]{inputenc}
\usepackage{graphicx}
\usepackage{dcolumn}
\usepackage{bm}
\usepackage{bbding}
\usepackage{color}
\usepackage{amsmath}
\usepackage{mathptmx}
\definecolor{red}{rgb}{1,0,0}

\usepackage{epstopdf}
\begin{document}


\title{Ejecta, corolla and splashes from drop impacts on viscous fluids}
\author{Florence Marcotte}
\author{Guy-Jean Michon}
\author{Thomas S\'eon}
\affiliation{Sorbonne Universit\'e, CNRS, Institut Jean Le Rond d'Alembert, F-75005 Paris, France}
\author{Christophe Josserand}
\affiliation{Laboratoire d'Hydrodynamique (LadHyX), UMR7646 CNRS-Ecole Polytechnique, 91128 Palaiseau CEDEX, France}
\begin{abstract}
{We investigate both experimentally and numerically the impact of liquid drops on deep pools of aqueous glycerol solutions with variable pool viscosity and air pressure. With this approach we are able to address drop impacts on substrates that continuously transition from low-viscosity liquids to almost solids. We show that the generic corolla spreading out from the impact point consists of two distinct sheets, namely an ejecta sheet fed by the drop liquid and a second sheet fed by the substrate liquid, which evolve on separated timescales. These two sheets contribute to a varying extent to the corolla overall dynamics and splashing, depending in particular on the viscosity ratio between the two liquids.
}
\end{abstract}

\maketitle

\begin{figure}[h]
\centering
\noindent\includegraphics[width=1\hsize]{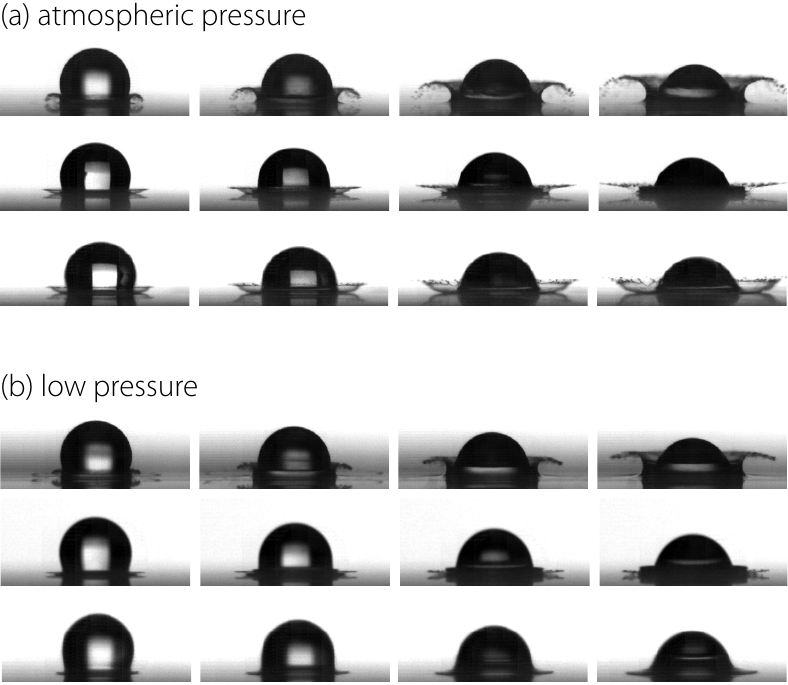}
\caption{Time sequences of the impacts of an ethanol drop for ${{\rm Re}= 5840}$ and ${{\rm We}=1080}$ on a viscous liquid pool with a viscosity ratio $\beta=0.95, \ 94$ and $\ 812$ (from top to bottom): (a) at atmospheric pressure; (b) at a lower pressure of one fourth of the atmospheric pressure $P=24.8 \, {\rm kPa}$. The snapshots are shown at $120, \, 180, \, 240$ and  $300 \,  \mu$s  (from left to right) after impact time.
}
\label{fig1}
\end{figure}

\begin{figure}
\centering
\noindent\includegraphics[width=1\hsize]{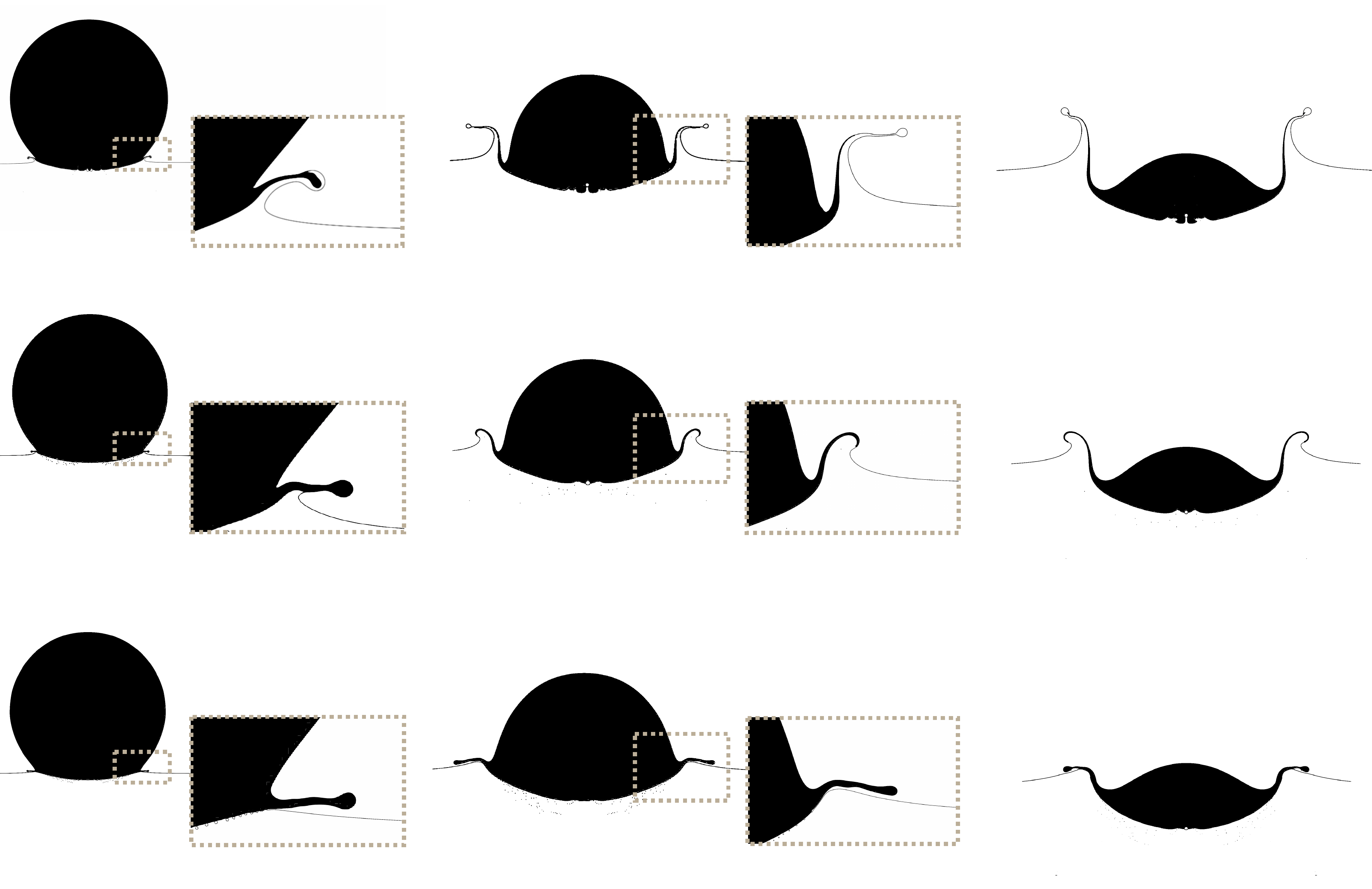}
\caption{Numerical time sequences with blowups on the interface region for $\beta=5$ (top), $\beta=50$ (middle) and $\beta=200$ (bottom), shown from left to right at dimensionless times $t=\{ 0.375,1, 2\}$ respectively (impact time is $t=0.2$). Here $Re=6000$ and $We=440$. For the $\beta=5$ case the secondary droplets detaching from the drop liquid were removed after their emission throughout the simulation.}
\label{sim-imp}
\end{figure}

\begin{figure}[h]
\centering
\noindent\includegraphics[width=1\hsize,trim=0 10 0 50,clip=true]{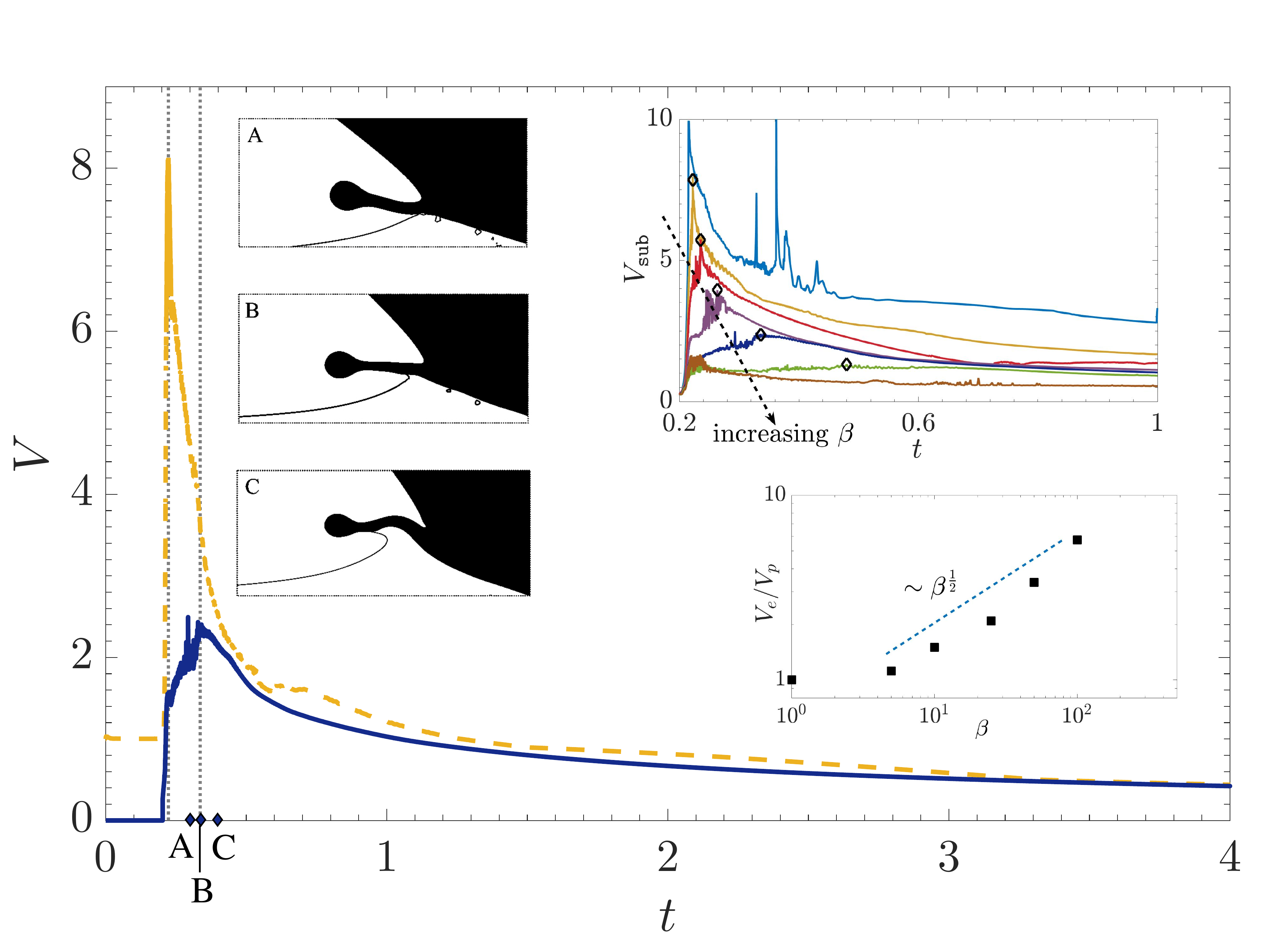}\\
\caption{Time-evolution of the maximal velocity monitored in the drop liquid (dashed line) and the substrate liquid (plain line) in a simulation with $\beta=50$, $Re=6000$ and $We=440$. The three snapshots \textit{(first inset)} show blowups on the interface region at times $A$, $B$, and $C$. \textit{Second inset:} Maximal velocity in the substrate liquid for $\beta$ in the $1-200$ range. The peak corresponding to the emission of the substrate jet (wherever defined) is marked by an empty diamond. \textit{Third inset:} $V_e/V_p$ (where $V_e$ is the peak velocity in the drop liquid and $V_p$ in the pool liquid) as a function of $\beta$ (full, black squares).}
\label{defo}
\end{figure}

Throwing a stone in a stagnant pond or letting a waterdrop fall onto a dry plate equally contribute to the \textit{active pleasures of water splashing}~\cite{Pe81}, as does the rewarding observation of the short-lived liquid corolla which, in both cases, blooms on the impact point~\cite{Flower}. From a comprehensive point of view however, the dynamics of the two events remarkably differ -- if only for the matter-of-fact reason that the splashed liquid belongs to the projectile in the latter, and to the impacted substrate in the former. The case of a liquid drop hitting a liquid surface therefore raises a natural question: which of the two liquids feeds the corolla as it spreads out, develops and eventually disintegrates? How does the splashing dynamics relate to that of the two first problems?\\
Splashes are formally defined as the ejection of small droplets due to the large deformation of a liquid interface following an impact, and occur in a large diversity of problems related to challenging environmental and industrial applications ~\cite{Rein93,Yarin06,JTAR16,Gilet15,liu93,Panao10,Ink-Mat10}.
In particular, two manifestations of splashing are discussed at length in the literature, referred to as `prompt splash'~\cite{tho02,Latka12,deegan08} and `crown splash'~\cite{Rioboo03,deegan08}, and mostly discriminated by the dynamics, shape and behaviour of the liquid sheet (which we will generically refer to as \textit{corolla}) whose desintegration results in the ejection of droplets. Prompt splash is associated with the early destabilisation of a thin ejecta sheet shooting out almost horizontally from the impact point: this axisymmetric liquid jet expands radially, bends upwards and disintegrates into small and fast droplets~\cite{tho02,Xu05,deegan08,Driscoll11}. On the other hand, crown splash originates in the destabilisation of an almost vertically expanding liquid sheet (sometimes referred to as Peregrine sheet~\cite{deegan08}) rising out of the impact region~\cite{Pe81,Edgerton,deegan08}. Here the `crown' emerges through the fingering of the liquid rim at the leading edge of the Peregrine sheet, owing to coupled Rayleigh-Taylor (RT) and Rayleigh-Plateau (RP) instabilities, and produces somewhat larger droplets~\cite{Edgerton,deegan08,DeZh10,Gilou14}.\\
However, some considerable confusion remains regarding the precise characterisation of these two splashing regimes and associated corolla dynamics. Indeed, the complicated splashing phenomenology rarely allows for such a clear separation between the corollas prone to `prompt' or `crown' splash ~\cite{deegan08,wang00}: for instance, splashing corollas consisting of mingled Peregrine and ejecta sheets have been identified for drop impact experiments on thin liquid layers~\cite{deegan08,Thoraval12}. Nevertheless, the nature of the impacted body (whether solid or liquid) seems to have at least a discriminatory effect on the corolla (and thus the splashing) dynamics: impacts on solid surfaces favour the development of an ejecta sheet-corolla~\cite{wang00,Rio01,JZ03,Xu05,Xu07}, whereas Peregrine sheets-corollas are observed for impacts on liquid pools or layers~\cite{Deegan12,Deegan12B,Gilou14}. This view is further suggested when decreasing the surrounding gas pressure, which can eventually suppress splashing on smooth solid substrates~\cite{Xu05,Latka12} whereas impacts on liquid films appear almost unchanged~\cite{Sussman16,Thor11}.\\
This apparent distinction -- between prompt- and crown-splashing corollas on the one hand, between solid and liquid surfaces on the other hand -- motivates the present letter, where impacts on a smooth solid or same-liquid body are viewed as asymptotic cases of the same generic problem, namely that of a liquid drop impacting a viscous liquid of variable viscosity~\cite{Rois18}. We aim at determining how the corolla structure and evolution depend on the substrate state and reconciling the observations made for impacts on solids and liquids. For that purpose, we combine an experimental study of the impact and splashing of ethanol drops on deep pools of aqueous glycerol solutions at fixed impact velocity and variable air pressure, with a numerical study of the corolla structure in a simple, axisymmetric impact model with varying substrate viscosity. Our approach allows for a continuous transition from impacts on liquid pools to impacts on (almost) solid substrates and provides a unified framework for understanding the mechanisms beneath corolla formation and splashing.\\
Ethanol droplets of diameter ${D=2.62\pm 0.11~{\rm mm}}$, released from a nozzle located at height 
$H=60 ~{\rm cm}$, impact a deep liquid pool of glycerol/water solution (tank dimensions: $80 \times 80 \times 50 ~{\rm mm^3}$) at velocity $U_0 = 3.39\pm 0.17 ~{\rm m}\cdot{\rm s}^{-1}$. The dynamic viscosities of ethanol, water and glycerol at $20^\circ~\rm C$ are respectively $\mu_e=0.0012$, $\mu_w=0.001$ and 
$\mu_o=1.49~{\rm kg/m/s}$. The viscosity of the glycerol/water solution $\mu_p$ ranges from $\mu_w$ to 
$\mu_o$ so that its ratio to ethanol viscosity ${\beta= \mu_p/\mu_e}$ varies from $0.95$ to $1000$. These three fluids are miscible, with the respective densities 
$\rho_e=789$, $\rho_w=1000$ and $\rho_o=1260~{\rm kg/m^3}$, and the air-liquid surface tensions $\gamma_e=0.022$, $\gamma_w= 0.072$ and $\gamma_o=0.064~\rm kg/s^2$. The experiments are 
performed in a closed chamber connected to a vacuum pump where the pressure could be lowered down to 
$8~{\rm kPa}$. Impacts were recorded using a high-speed camera Photron SA-5. 
Since the drops diameter and impact velocity were fixed throughout the experiments, the problem is characterised by fixed Reynolds number ${\rm Re}= \rho_e U_0 D/\mu_e=5840$ and Weber number ${\rm We}= \rho_e U_0^2 D/\gamma_e=1080$. The other dimensionless parameters involved here are the various density, viscosity and surface tension ratios,
although in the present study only the drop/pool viscosity ratio $\beta$ and the air/ethanol density ratio $\tilde{\alpha}=\rho_g/\rho_e$ were effectively varied, the other ones being either fixed or varying in a much less significant amount. Because of Maxwell's law on gas viscosity the air/ethanol viscosity ratio ${\tilde{\beta}=\mu_g/\mu_e}$ does not vary with the air pressure. 
Finally, the large Péclet number ${\rm Pe}=U_0D/\kappa \sim 10^3$ (with $\kappa$ the molecular diffusion coefficient of ethanol in water) implies that mixing operates on timescales larger than the typical duration of the experiments.\\
Figure \ref{fig1}a illustrates the evolution of the corolla dynamics as the liquid pool viscosity increases at 
(constant) ambient pressure. Although splashing occurs in all cases, striking differences firstly lie in the corolla shape, which transitions from a 
downwards-curved corolla at low $\beta$ (low-viscosity substrate, top sequence) to a upwards-curved corolla at $\beta=812$ (high-viscosity substrate, bottom sequence). Based on the early expansion of the axisymmetric jet that spreads out from underneath the drop, it is tempting to identify the high-$\beta$ corolla with a typical ejecta sheet, which first expands horizontally before it is deflected upwards, and the low-$\beta$ corolla with a typical Peregrine sheet, which springs up almost vertically despite a downwards bending of the leading edge. What happens in between ($\beta=94$, middle sequence) is definitely more ambiguous: whereas the early jet evolution reminds of that of the ejecta sheet, in the later stages the corolla almost appears as a purely radially expanding sheet laying on the top of a rising pedestal.\\
The pressure in the surrounding air was decreased in the experiments to further reveal the discrepancy in dynamics between  the low- and high-$\beta$ corollas. Lowering the air pressure 
for the highest substrate viscosities weakens and eventually suppresses the splash, as illustrated in figure \ref{fig1}b (bottom sequence), consistently with observations for impacts on solid substrates~\cite{Xu05}. The decrease in air pressure tends to stabilise the crawling ejecta sheet until it cannot detach from the substrate. 
Conversely, for the lowest substrate viscosity investigated here ({$\beta = 0.95$}, top sequence in figure \ref{fig1}b), lowering the air pressure cannot suppress the formation of the Peregrine sheet, whose destabilisation eventually yields splashing. In fact, we observe the corolla shape to experience the same overall evolution whereas the decrease in air pressure tends to inhibit the early 
disintegration of the leading edge.\\
Importantly, the middle sequence in figure \ref{fig1}b shows that for intermediate substrate viscosity ({$\beta=94$}), some of the corolla features still respond to the pressure decrease: even though splashing eventually occurs, the destabilisation of the nearly horizontal jet, the ejecta sheet, is inhibited at low pressure. Additionally, the lift-up of this first sheet is significantly reduced, although to a smaller extent than in the high-$\beta$ case - but here a kinematic deviation is also induced by the formation of a crater. On the other hand the pressure decrease does not affect the dynamics of the pedestal, which forms slightly after the emission of the ejecta sheet (note the dark bump spreading on the last snapshots of figures \ref{fig1}a and \ref{fig1}b, middle sequence) and does not splash. Our results therefore suggest that the transition between low- and high-$\beta$ impact regimes, which affects both the corolla dynamics and its sensitivity to pressure, can be interpreted in terms of two coexisting structures: the ejecta sheet shooting out at early times, and a weaker, slower sheet (as $\beta$ increases) corresponding to the `pedestal'.\\
Numerical simulations were carried out using the {\sc Gerris} flow solver~\cite{GerrisVOF,Pop03} to understand the nature and role of these two structures in the transition between the low- and high-$\beta$ regimes, at ambient pressure and in a simplified configuration where the drop and pool liquids only differ in viscosity. {\sc Gerris} has been extensively used to solve the incompressible, two-phase Navier-Stokes (NS) equations in a wide range of multiphase problems \cite{gerrisbiblio} and has been validated against various experiments, ranging from drop impacts \cite{Gilou14,Gilou15} to busting bubbles \cite{deike18} or breaking waves \cite{deike17}. The gaseous and liquid phases were discriminated using a characteristic function $\chi_1$ ($\chi_1=1$ in the liquid and $\chi_1=0$ in the air), associated with a surface tension which we assumed to be constant; for simplicity $\gamma=\gamma_w$ and $\rho=\rho_p$ in all the liquid phase. The viscosity jump between the (slowly) miscible drop and pool liquids was described by means of a second characteristic function $\chi_2$ simply defined here as a non-diffusive tracer, such that  $\chi_2=1$ in the drop liquid and $\chi_2=0$ otherwise, and without associated surface tension. Both $\chi_1$ and $\chi_2$ were advected using a VOF method.\\
The density and viscosity fields were defined as:
$\rho(\textbf{x},t) = \rho_p \chi_1 + \rho_g (1-\chi_1)$ and 
$\mu(\textbf{x},t) = \mu_p \chi_1 (1-\chi_2)+\mu_e \chi_1 \chi_2+ \mu_g (1-\chi_1)$, and
the NS equations were solved numerically using adaptive mesh refinement in axisymmetric geometry. The computational domain is a square box of dimensionless size $L=4$, where the unit length is the initial radius of liquid droplet (released at height $H=0.2$ with dimensionless velocity $U_0=1$), and $d=1.7$ is the liquid pool depth.
The prescribed boundary conditions are no-slip on the bottom boundary, axial symmetry on one side, and free outflow otherwise. 
Importantly, the prescribed axial symmetry filters out the inherently three-dimensional mechanisms for droplet ejection, so that unlike our experiments the simulations do not (and indeed cannot) address splashing but rather the transition in corolla structure from low- to high-$\beta$ regimes. Also, because of the simplifications made in our numerical model (in particular a unique liquid/air surface tension), the control parameters were chosen so as to achieve qualitatively similar impact behaviors while considering smaller (and numerically less demanding) $We$ numbers, corresponding to a surface tension closer to that of water than ethanol: here $Re=6000$, $We=440$, $\tilde{\alpha}=0.0015$ and $\tilde{\beta}=0.015$, for which the transition was observable from the $\beta \in \{1-200\}$ range. (For example, these impact parameters also describe a waterdrop of diameter $1.1~{\rm mm}$ impacting a surface at speed $5.3~{\rm m/s}$, a situation close to the relevant regime for irrigation sprinklers in agriculture \cite{kincaid} or inkjet printing \cite{Ink-Mat10}.) The adaptive quadtree grid was refined up to 12 levels of refinement, which convergence tests proved to be sufficient.\\
Figure \ref{sim-imp} shows sequences of simulations snapshots (with blowups on the interface) for impacts on increasingly viscous substrates. 
Here the dark areas correspond to the liquid from the impacting drop and the liquid-air interface is highlighted by the thick line, so that the contribution of both the drop and the pool liquids to the corolla structures could be monitored in time. At low $\beta$ (upper sequence, $\beta=5$), a jet consisting of both the impacting fluid and the substrate is emitted and forms a downwards-curved, Peregrine-like corolla as it expands both vertically and radially, until the corolla leading edge is pulled upright by capillary forces. A similar evolution was observed for drop impacts on shallow pools of the same liquid~\cite{Ray16}. At intermediate $\beta$ (middle sequence, $\beta=50$), a first jet consisting solely of the drop liquid is emitted almost horizontally at early times, and is caught up at later times by a weaker sheet induced by the substrate deformation. The time separation between the early emission of this ejecta sheet (from the drop) and the slower formation of the substrate sheet becomes clearer as $\beta$ further increases (bottom sequence, $\beta=200$). Eventually, the two sheets merge due to capillary forces, giving rise to a single structure that becomes weaker with increasing $\beta$.\\
Our numerical results suggest a new scenario shedding lights on the experimental results: {\it two} jets are always generated in the impact, one emitted from the drop - and feeding what would in fact 
appear as an unequivocal ejecta sheet at sufficiently high $\beta$ - and the other one from the liquid substrate, the pedestal - feeding what would appear as an unequivocal Peregrine sheet at sufficiently low $\beta$. When the viscosity ratio $\beta$ is weak, these jets form almost simultaneously and then rapidly merge so that a single sheet seems to develop at short times, with an initial angle of $45^o$ with the horizontal. The strong vertical expansion of the resulting two-fluid corolla is mostly driven by the strong substrate deformation, and the overall dynamics is that of a typical Peregrine sheet. As $\beta$ increases, the substrate jet is delayed and becomes weaker, so that observations show a single-fluid ejecta sheet emitted almost horizontally, caught up at later times by the substrate sheet. \\
Figure \ref{defo} shows the time-evolution of the maximal velocity monitored in both liquids for the intermediate case $\beta=50$: the solid and dashed lines correspond to its largest values in the pool and the drop liquids respectively. As observed in~\cite{Ray16} for a single liquid, the maximal velocity displays a peak at the time where the ejecta jet is generated. Here two peaks are observed successively, first in the drop liquid at impact time (${t \sim 0.2}$; first vertical, dotted line), then in the pool liquid (${t \sim 0.33}$ for $\beta=50$; second vertical, dotted line). The three simulations snapshots in the first inset in Figure \ref{defo} are blowups on the interface region at the time where the second, weaker peak is reached (${t=B}$), shortly before (${t=A}$) and shortly after (${t=C}$): this 
peak corresponds to the formation of a second jet issued from the substrate, which catches up on the first ejecta sheet at later times. While the ejecta sheet always shoots out immediately after impact time and with the same typical velocity, the second inset in Figure \ref{defo} shows the evolution of the maximal velocity in the pool liquid for increasing $\beta$: the velocity peak (marked by empty diamonds) corresponding to the emission of the substrate jet is delayed and its amplitude decays as the pool viscosity increases, until the well-defined peak dissolves into a smooth bump and eventually vanishes for $\beta=200$. The last inset in Figure \ref{defo} shows the evolution of the ratio between the two peak velocities ($V_e/V_p$, with $V_e$ the peak velocity in the drop liquid and $V_p$ in the pool liquid) as a function of $\beta$ (full, black squares), wherever the two peaks are easily defined. This evolution is consistent with the self-similar theory developed by~\cite{JZ03} for drop impacts on thin layers of the same liquid. Their analysis predicts that the velocity of the jet generated by the impact should scale like $\sqrt{{\rm Re}} \, U_0$, where ${\rm Re}$ is the Reynolds number based on the liquid viscosity, showing good agreement with experiments~\cite{tho02} and numerical simulations~\cite{Ray16}. This theoretical prediction can be transposed to the present case ($\beta \neq 1$) by conjecturing that the typical velocities of the two distinct liquid jets respectively scale like $\sqrt{{\rm Re}} \, U_0$ (for the drop jet) and $\sqrt{{\rm Re}/\beta} \, U_0$ (for the substrate jet): indeed the prediction is found to provide a good approximation of the trend observed in figure \ref{defo} (last inset).\\
By addressing the problem of a drop impact on a liquid substrate with highly variable viscosity, the present study reconciles the observations of the very diverse corolla shapes and splashing behaviors generated by impacts on solid or liquid surfaces. Our results show that the transition between the impact-on-liquids and impact-on-solids regimes is a continuous one, and that the liquid corolla spreading out from the impact region generically consists of two sheets, respectively fed by the drop and the substrate jets, which as the substrate viscosity increases form on increasingly separated timescales. 
At low $\beta$, the drop and substrate sheets immediately merge into a single Peregrine sheet. As $\beta$ increases, the substrate sheet progressively dissolves into a mild surface wrinkle and vanishes at large $\beta$, its weakening resulting in the suppression of (crown) splashing from the substrate liquid. Our results allow for new and more consistent definitions of the ejecta sheet as the (possibly short-lived) jet of drop liquid \textit{before} its merging with the substrate jet, and of the prompt splash as that of the ejecta (unmerged as yet). As opposed to the substrate sheet, and presumably because of its weak emission angle, this ejecta is highly sensitive to the ambient air pressure and prone to prompt splash as long as it can detach from the substrate. The (mildly) stabilising effect of low gas pressure on the Peregrine sheet leading edge at low $\beta$ can then be explained by the suppression of the early perturbation induced by the prompt splash.
Our results suggest that the `splashing number' used in many different impact contexts to characterise the splashing threshold (see \cite{StHa81,MST95,YW95,Yarin06,JTAR16} and references herein) should be revisited in the light of the corolla two-sheet structure. Different thresholds could be introduced depending on the nature of the sheet driving the dynamics, consistently with recent observations related to crown splashes~\cite{Bang18}. Importantly, our results indicate that the two jets relative dynamics could also modify the splashing threshold as $\beta$ increases, due to their varying tangential velocity (as shown by \cite{Bird09} for drop impacts on moving solid substrates), or the inhibition by viscosity of the destabilising von K\'arman vortex street observed at the drop/substrate interface for $\beta=1$~\cite{Thoraval12}. Finally, it would be interesting to address the situation opposite to the one we have investigated here, namely the impact of drops with highly variable viscosity on a liquid substrate.

{\it Acknowledgement.} The authors wish to thank St\'ephane Popinet and Pascal Ray for helpful discussions, and two anonymous referees for their very constructive comments.

\bibliography{goutte}

\end{document}